\documentclass[a4paper,11pt]{article}

\usepackage{pos}

\title{IceCube search for neutrinos from GRB 221009A}

\ShortTitle{IceCube search for neutrinos from GRB 221009A}

\author{The IceCube Collaboration \\{\normalsize \normalfont(a complete list of authors can be found at the end of the proceedings)}\\}

\emailAdd{karlijn.kruiswijk@uclouvain.be}
\emailAdd{bbrinson9@gatech.edu}
\emailAdd{rprocter@umd.edu}
\emailAdd{jthwaites@icecube.wisc.edu}
\emailAdd{Nora.valtonen-mattila@icecube.wisc.edu}

\abstract{

GRB 221009A is the brightest Gamma Ray Burst (GRB) ever observed. The observed extremely high flux of high and very-high-energy photons provide a unique opportunity to probe the predicted neutrino counterpart to the electromagnetic emission. We have used a variety of methods to search for neutrinos in coincidence with the GRB over several time windows during the precursor, prompt and afterglow phases of the GRB. MeV scale neutrinos are studied using photo-multiplier rate scalers which are normally used to search for galactic core-collapse supernovae neutrinos. GeV neutrinos are searched starting with DeepCore triggers. These events don't have directional localization, but instead can indicate an excess in the rate of events. 10 GeV - 1 TeV and >TeV neutrinos are searched using traditional neutrino point source methods which take into account the direction and time of events with DeepCore and the entire IceCube detector respectively. The >TeV results include both a fast-response analysis conducted by IceCube in real-time with time windows of T$_0 - 1$ to T$_0 + 2$ hours and T$_0 \pm 1$ day around the time of GRB 221009A, as well as an offline analysis with 3 new time windows up to a time window of T$_0 - 1$ to T$_0 + 14$ days, the longest time period we consider. The combination of observations by IceCube covers 9 orders of magnitude in neutrino energy, from MeV to PeV, placing upper limits across the range for predicted neutrino emission.

\vspace{4mm}
{\bfseries Corresponding authors:}
Karlijn Kruiswijk$^{1*}$, Bennett Brinson$^{2}$, Rachel Procter-Murphy$^{3}$, Jessie Thwaites$^{4}$, Nora Valtonen-Mattila$^{5}$,  \\
{$^{1}$ \itshape Centre for Cosmology, Particle Physics and Phenomenology - CP3, Universit{\'e} catholique de Louvain, Louvain-la-Neuve, Belgium}\\
{$^{2}$ \itshape School of Physics and Center for Relativistic Astrophysics, Georgia Institute of Technology}\\
{$^{3}$ \itshape Dept. of Physics, University of Maryland}\\
{$^{4}$ \itshape Dept. of Physics and Wisconsin IceCube Particle Astrophysics Center, University of Wisconsin \textendash Madison}\\
{$^{5}$ \itshape Dept. of Physics and Astronomy, Uppsala University}\\[4mm]
$^*$ Presenter

\ConferenceLogo{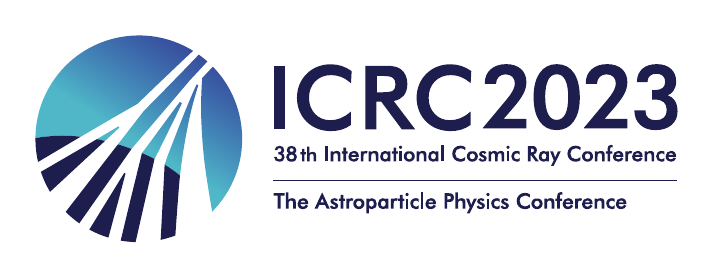}

\FullConference{The 38th International Cosmic Ray Conference (ICRC2023)\\ 26 July -- 3 August, 2023\\ Nagoya, Japan}
}

\begin{document}

\maketitle

\section{Properties of GRB221009A}\label{sec1}
At T$_0$=13:16:59.99 UT the Gamma-ray Burst Monitor (GBM) onboard the Fermi satellite was triggered on an extremely bright burst: GRB221009A. 
In realtime, the central 90\% containment of emission (T$_{90}$) was reported with a duration of $325.8\pm6.8$ s, beginning 221.1 s after T$_0$ \cite{GCN32642}. The burst was localized by the Swift Observatory at right ascension $\alpha=288.2645^\circ$, declination $\delta=+19.7735^\circ$, with a 90\% containment of 0.61" \cite{GCN32632}. The isotropic equivalent gamma-ray energy has been measured to be $\sim1.2\times10^{55}$ erg by Konus-Wind \cite{Frederiks:2023}.

\section{Neutrino searches with IceCube}\label{sec2}
IceCube is a cubic-kilometre neutrino detector buried in the glacial ice at the geographic South Pole \cite{Aartsen:2016nxy}. It consists of 5160 Digital Optical Modules (DOMs), each with a photomultiplier tube (PMT). These PMTs are able to observe the Cherenkov light emitted by relativistic charged particles created by neutrino interactions in the ice. In the center of IceCube, eight strings containing PMTs with a higher quantum efficiency are placed with a denser DOM and string spacing. This part of the detector is called IceCube-DeepCore, and has a reduced energy threshold down to about 0.5 GeV.

We perform four complementary IceCube analyses to search for neutrinos using the full energy range accessible using the IceCube detector: the Gamma-ray Follow-Up (GFU) sample for >100~GeV neutrinos \cite{IceCube:2016xci}, the GeV Reconstructed Events with Containment for Oscillations (GRECO) sample for $10 - 1000$~GeV neutrinos \cite{IceCube:2022lnv}, the Extremely Low Energy (ELOWEN) sample for $0.5 - 5$~GeV neutrinos \cite{IceCube:2021ddq}, and the MeV neutrino burst analysis for <1~GeV neutrinos \cite{IceCube:2011cwc}. The full explanation of these analyses from MeV to PeV can be found in  \citep{IceCube:2023rhf}.

\begin{figure}[htb]
    \centering
    \includegraphics[width = 0.65\textwidth]{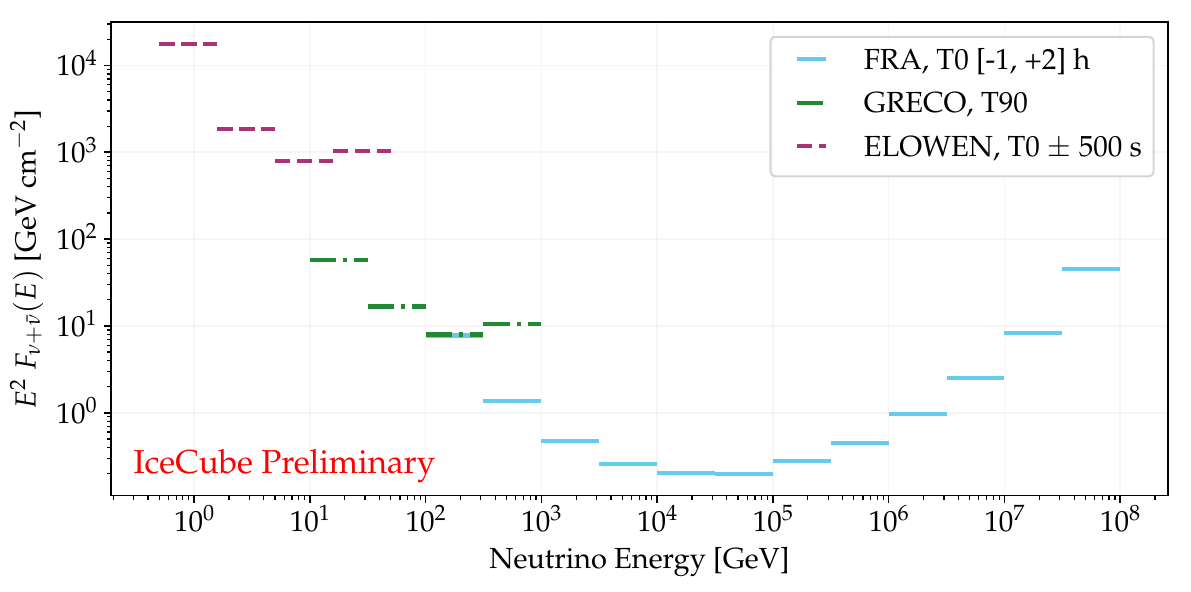}
    \caption{Time-integrated per-flavor $\nu+\bar{\nu}$ 90\% CL differential upper limits on the neutrino flux from GRB 221009A with the ELOWEN, GRECO, and GFU samples. We assume a time-integrated flux with a power-law spectrum $F_{\nu+\bar{\nu}}(E) \propto E^{-2}$ for the neutrino spectrum in each energy bin.}
    \label{fig:difflimits}
\end{figure}

\begin{figure}[htb]
    \centering
    \includegraphics[width = 0.65\textwidth]{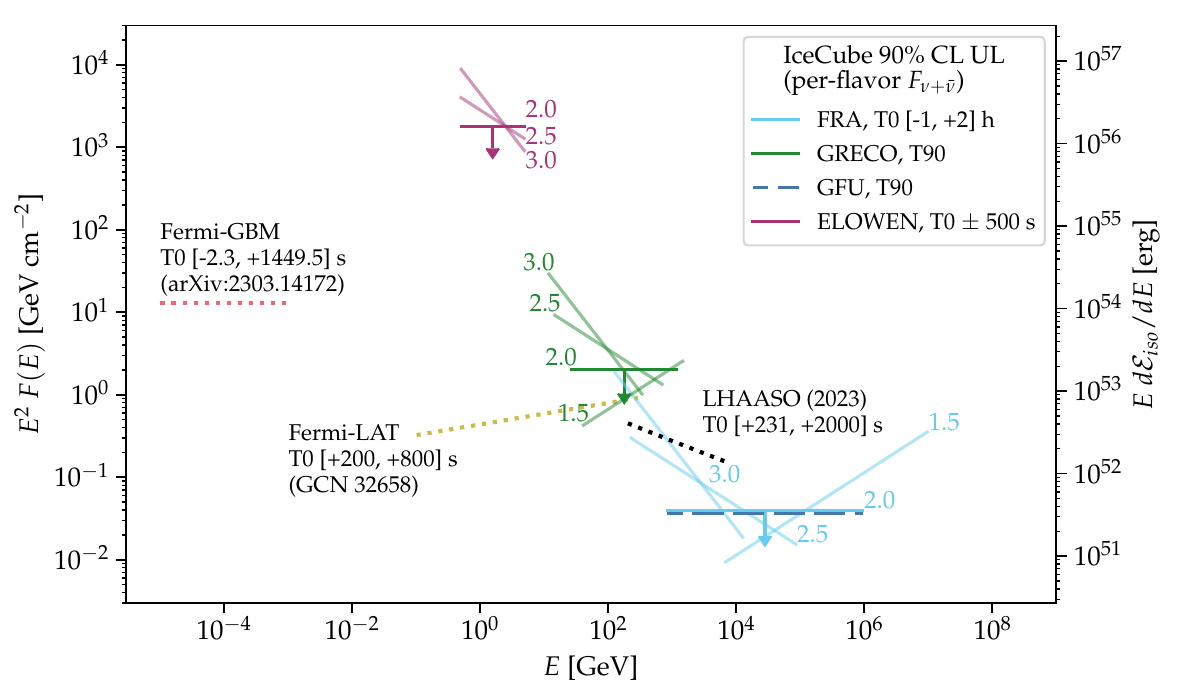}
    \caption{Gamma-ray observations and time-integrated upper limits on the neutrino flux of GRB 221009A. We show the gamma-ray observations reported by \cite[Table S2]{LHAASO_2023} in the brightest period of the emission. We show the total gamma-ray fluence observed by Fermi-GBM, as reported by \cite{FermiGBM:2023}, at $\gamma=2.0$ for visualization purposes. We also plot the gamma-ray observations from Fermi-LAT as reported in realtime \cite{GCN32658}, although this flux is known to be underestimated due to saturation effects. The right axis shows the differential isotropic equivalent energy, with $E^2 F_{\nu+\bar{\nu}}(E) = E d\mathcal{E}_{iso}/dE \times (1+z)/4\pi D_L^2$.}
    \label{fig:timeint_limits}
\end{figure}

No significant deviation from the background expectation was found in any of the analyses. Thus, we set upper limits on the neutrino flux from GRB221009A ($F(E) \equiv d^2N/dE/dA$) assuming fixed power law fluxes (GFU, GRECO, ELOWEN) or a generic blackbody spectrum (MeV). Differential limits of the GFU, GRECO and ELOWEN sample can be seen in Figure \ref{fig:difflimits}. We also show the time-integrated upper limits for each of these three data samples in Figure \ref{fig:timeint_limits}, as done in \cite{IceCube:2023rhf}, with updated gamma-ray observations from Fermi-GBM \cite{FermiGBM:2023} and LHAASO \cite{LHAASO_2023}. The MeV burst analysis upper limits are shown in Figure \ref{fig:MeV_dirac_delta}.

\section{Model constraints}\label{sec3}
In addition to the differential limits on the neutrino flux, constraints can be placed on models for neutrino emission from GRB~221009A in the in the relevant energy ranges for each data sample, as described in the following sections. 

\subsection{TeV Neutrino Emission Models}
Gamma ray bursts (GRBs) have long been thought to be a source of high energy neutrinos and ultra-high energy cosmic rays \cite{Waxman:1997}. In the fireball model, hot plasma is ejected from the GRB source, producing a "fireball" that is initially optically thick to radiation, but expands by radiation pressure until it becomes optically thin. The fireball engine produces internal shock collisions which accelerates baryons, producing ultra-high energy cosmic rays and neutrinos through photohadronic interactions. 

\begin{figure}[htb]
    \centering
        \centering
        \includegraphics[width=0.475\linewidth]{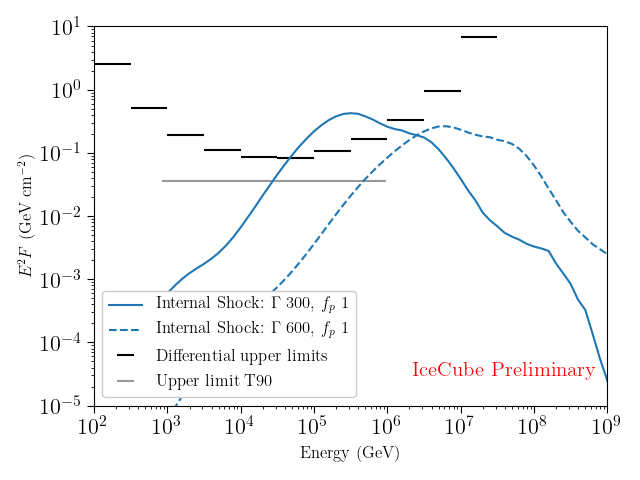}%
        \hfill
        \includegraphics[width=0.475\linewidth]{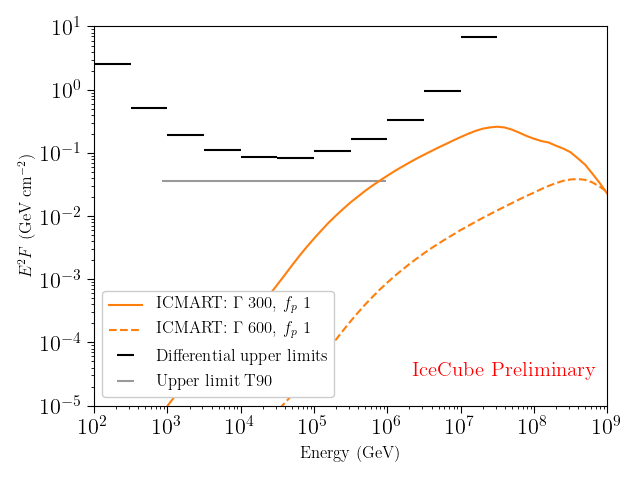}
    \caption{Neutrino spectra calculated for Lorentz factors of 300 and 600, for baryonic loading equal to 1, for the internal shock model (left) and ICMART (right) with the time-integrated 90\% CL differential upper limits on the neutrino flux from GRB221009A and 90\% CL upper limit from the Fast Response Analysis (GFU) sample in the T$_90$ time window.}
    \label{fig:TeVfluxSpectra}
\end{figure}

In this analysis, we investigate two different fireball models: the internal shock model \cite{Waxman:1997,Zhang:2013} and the Internal Collision-Induced Magnetic Reconnection and Turbulence (ICMART) model \cite{Zhang:2013,Zhang_2011}. The two models use similar mechanisms, but ICMART predicts neutrino production from magnetic reconnection at a larger radius from the source. In order to test these models, we generated neutrino spectra using \textit{Fireballet} \cite{Aartsen_2016} with varying baryon load and bulk Lorentz factors for the T$_90$ time window. These spectra can be seen in Figure \ref{fig:TeVfluxSpectra}.

 We assume that the proton spectrum follows a $E^{-2}$ power law. We used the redshift ($z=0.151$), isotropic-equivalent luminosity ($L_{\mathrm{iso}}=9.9\times 10^{53}$ erg/s), low-energy photon index ($\alpha=-1.583$), high-energy photon index ($\beta=-3.77$), and break energy ($E_{\mathrm{break}}=1387$ keV) from the best-fit parameters in Fermi-GBM \cite{FermiGBM:2023}. The emission radius of neutrinos from internal shocks scales with the time variability ($t_{\mathrm{var}}$) \cite{Waxman:1997}. We use a time variability of 0.1~s to match the model used by GBM \cite{FermiGBM:2023}. For the ICMART model, we follow the procedure from \cite{Aartsen_2016} and assume $t_{\mathrm{var}}=1$~s because the ICMART model has a larger radius. For both models, the high-energy photon index, low-energy photon index, and break energy was calculated from the lightcurve between $277-323$~s from GBM \cite{FermiGBM:2023}. This time interval was chosen because Fermi-GBM did not have data issues within the time interval and the time interval falls within the T$_{90}$ time window we are testing. This flux was then injected to the previous analysis \cite{IceCube:2023rhf} in order to calculate constraints on the internal shock and ICMART models. 

\begin{figure}[htb]
    \centering
        \centering
        \includegraphics[width=0.475\linewidth]{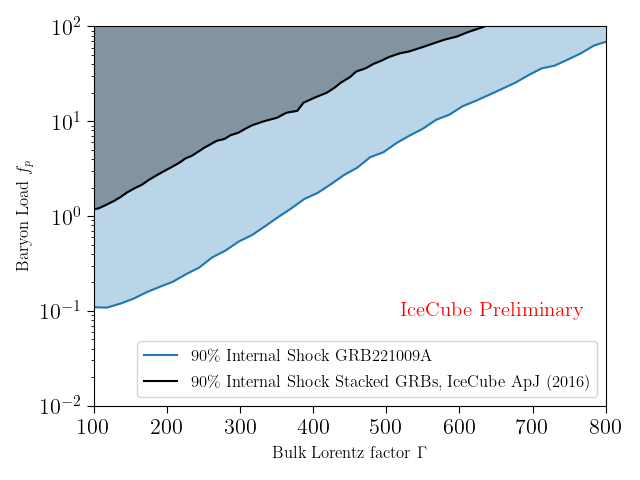}%
        \hfill
        \includegraphics[width=0.475\linewidth]{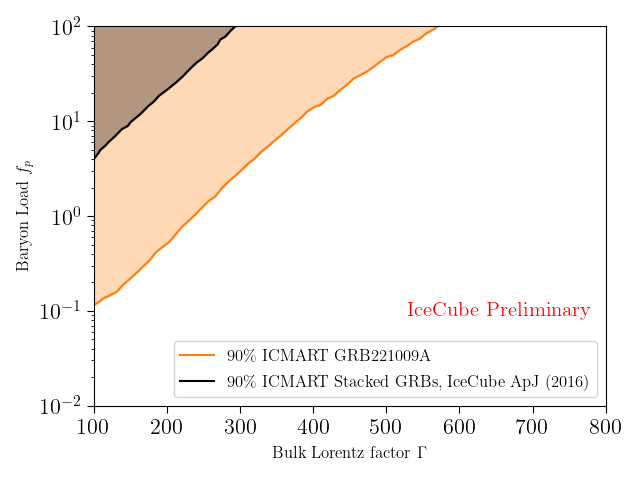
        }
    \caption{The 90\% CL on the Lorentz factor and baryon load for GRB221009A compared to previously published \cite{Aartsen_2016} upper limits for the internal shock model (left) and ICMART (right). For the internal shock model, $t_{vars} = 0.1$ for this work and $t_{vars} = 0.01$ for previous limits. The shaded regions are excluded by their respective limits.}
    \label{fig:TeVExclusion}
\end{figure}

We calculated the 90\% upper limits for various Lorentz factors and baryon loads for the internal shock model and ICMART and compared to previous limits set by IceCube \cite{Aartsen_2016}. Our results are summarised in Figure \ref{fig:TeVExclusion}.

The previous IceCube limit stacked several hundred GRBs and assumed average GRB parameters while the limits presented here are specific to GRB221009A. The limits set by GRB221009A are more constraining in all regions in the internal shock model and ICMART model, but Fermi-GBM set a lower limit on the Lorentz factor of $\Gamma \geq 780$ \cite{FermiGBM:2023}, with which we can exclude at a baryon load of 60.6. 

\subsection{GeV Neutrino Emission Models}
Independent of the production of high-energy cosmic rays, it is still possible to produce neutrinos in the GeV range. The GRB jet is expected to contain coupled outflows of protons and neutrons that accelerate within the jet. However, once the \textit{n-p} scattering time $t_{np}\sim1/(n_p\sigma_{np}c)$ becomes longer than the fireball's expansion time $t_{\mathrm{exp}}\sim r/c\Gamma$, the neutrons and protons will begin to decouple. The protons will continue to accelerate, but the neutrons will not, increasing their relative velocity until inelastic \textit{n-p} collisions occur. These collisions will produce pions capable of escaping the jet and decaying to produce neutrinos \cite{Bahcall:2000}. Neutrino emission from GRB221009A via this process (hereafter referred to as the quasi-thermal decoupling model) is predicted to be in the $\sim1-10$ GeV range \cite{Murase:2022}.

In addition to neutrinos from \textit{n-p} decoupling in each outflow, neutrino production is expected due to collisions between outflows. After decoupling, a given neutron outflow will cease acceleration. However, as subsequent proton outflows continue to accelerate, they will catch up to and collide with these free-streaming neutrons, leading to further \textit{n-p} collisions and further neutrino production \cite{Beloborodov:2009be}. Neutrino emission from GRB221009A via this process (hereafter referred to as the quasi-thermal collision model) is predicted to be in the $\sim10-100$~GeV range.

In both cases, the expected neutrino energy fluence is proportional to the kinetic energy of the proton outflow, $\xi_N\mathcal{E}_{\gamma}^{\mathrm{iso}}$, where $\mathcal{E}_{\gamma}^{\mathrm{iso}}$ is the isotropic equivalent gamma-ray energy of the GRB and $\xi_N$ is the nucleon loading factor. Of course, both models also depend on the Lorentz factor $\Gamma$, which can be between 100 and 1000. 
Additionally, the decoupling model depends on $\zeta_n$, the neutron-to-proton ratio, while the collision model depends on $\tau_{np}$, the \textit{n-p} optical depth  \cite{Murase:2022}. As the value of $\mathcal{E}_{\gamma}^{\mathrm{iso}}$ for GRB221009A has been measured, this allows us to place constraints on $\zeta_n\xi_N$ via the decoupling model, or $\tau_{np}\xi_N$ via the collision model.

To test these models, we utilized the previous upper limits set by IceCube on neutrino emission from GRB221009A in the range between 1 - 100 GeV. For the decoupling model, we used the ELOWEN upper limits, and for the collision model, we used the GRECO-Astronomy upper limits.

\begin{figure}
    \centering
    \includegraphics[width= 0.485 \textwidth]{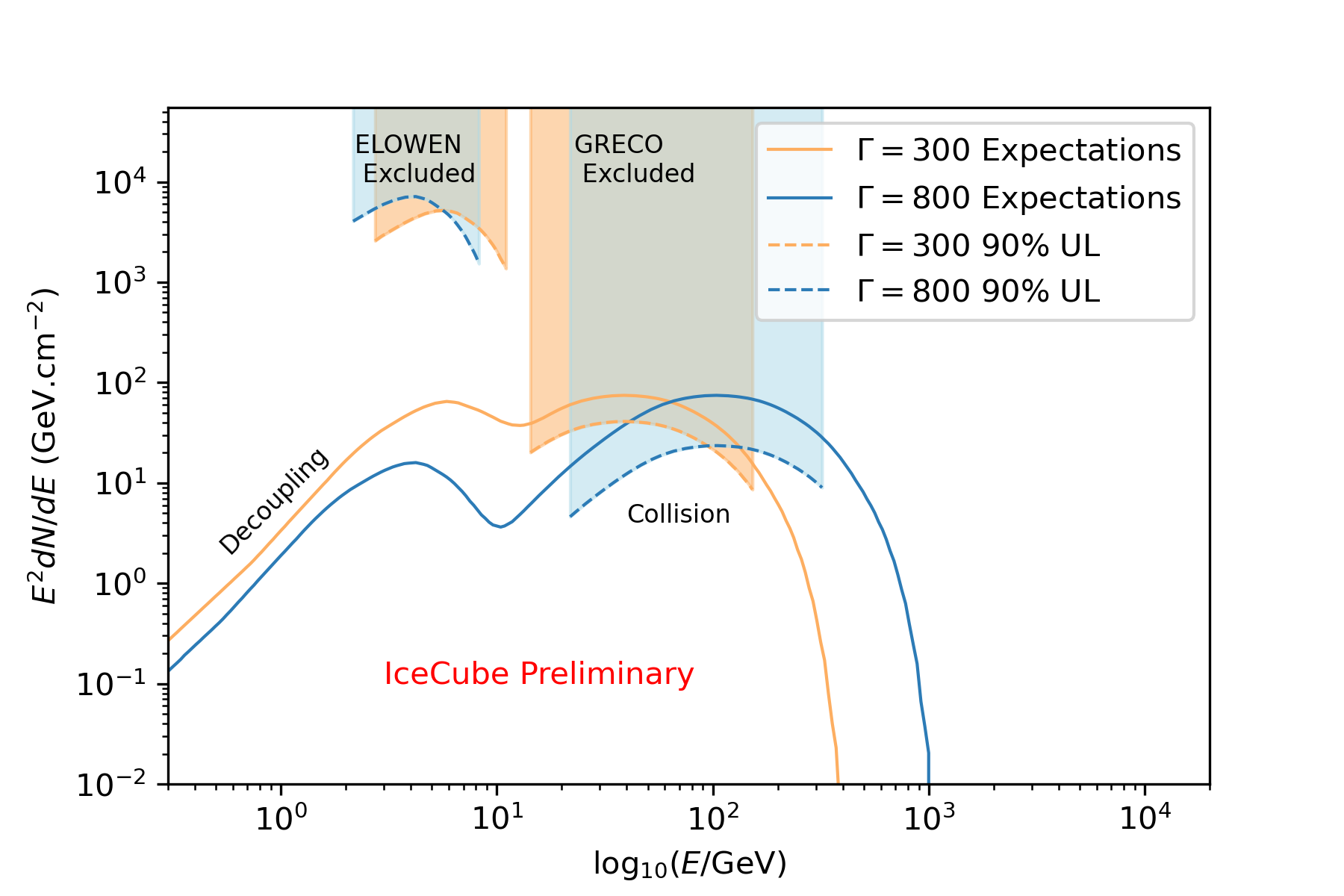}
    \hfill
    \includegraphics[width = 0.485 \textwidth]{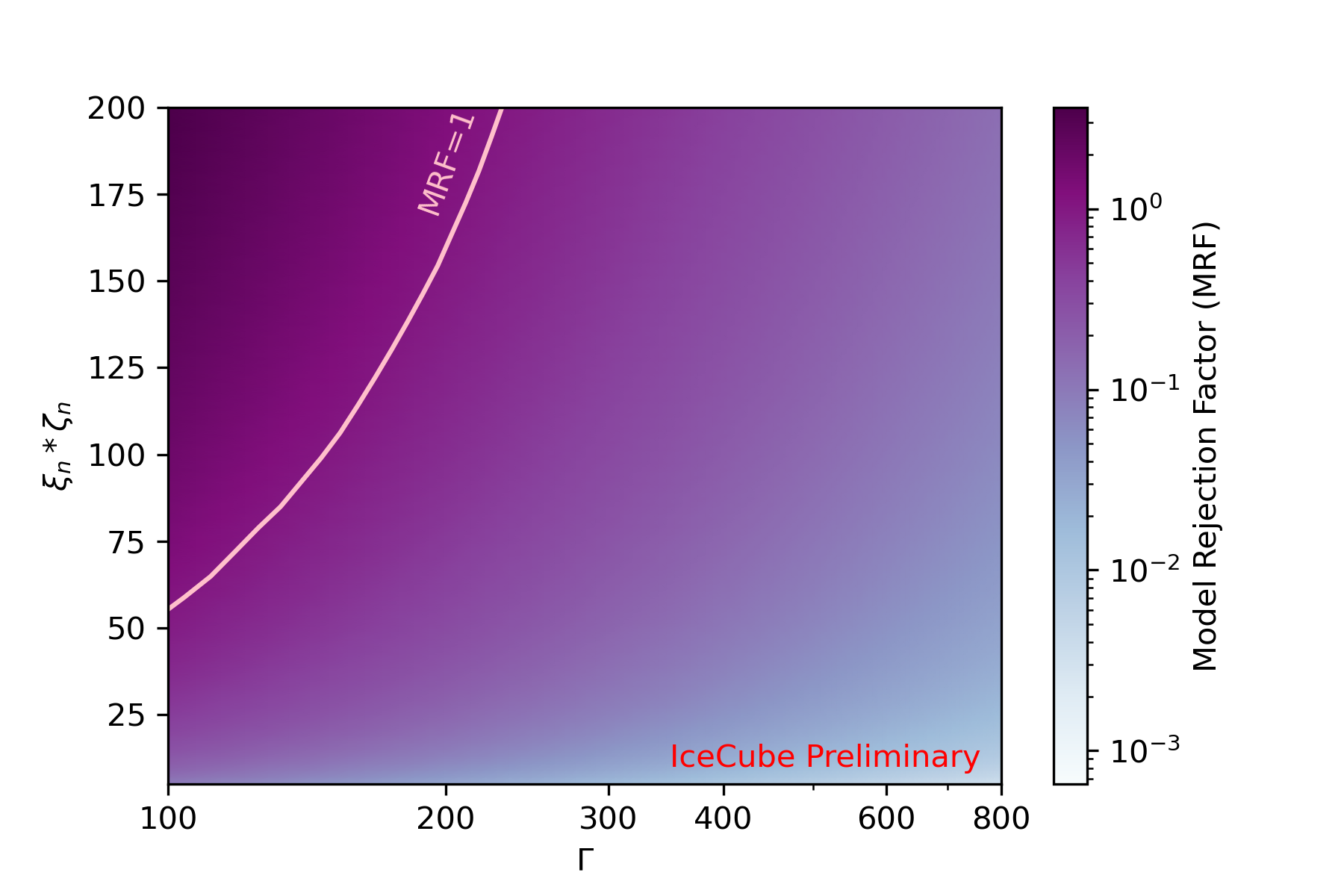}
    \caption{(left) The collision and decoupling models where $\zeta_n\xi_N=5$ for the decoupling model and $\tau_{np}\xi_N=5$ for the collision model. The 90\% upper limits are given using the 2200~s time window.
    (right) The model rejection factor (MRF) for the decoupling model calculated with the upper limits from ELOWEN for the 1000~s time window.}
    \label{fig:GeVModel}
\end{figure}

For each model, we used the neutrino flux as calculated by Murase et al \cite{Murase:2022}, normalised to $\zeta_n\xi_N=5$ (decoupling), $\tau_{np}\xi_N=5$ (collision), $\mathcal{E}_{\gamma}^{iso}=1.2\times10^{55}$ erg, and various choices of $\Gamma$. By injecting neutrinos into our dataset according to this flux, we were able to determine $N_{\mathrm{lim}}$, the number of neutrinos compatible with the previous IceCube limits. We then computed $N_{\mathrm{exp}}$, the expected number of neutrinos from GRB221009A, by integrating this flux over the effective area of the respective dataset at the location of this GRB. The ratio $\frac{N_{\mathrm{lim}}}{N_{\mathrm{exp}}}$ then gives the fraction of the model flux that is consistent with our non-detection of neutrinos from GRB221009A, the inverse of which is used as the Model Rejection Factor (MRF). The MRF can be used to place constraints on the value of $\zeta_n\xi_N$ in the decoupling model, or $\tau_{np}\xi_N$ in the collision model. The expected neutrino flux is shown in Figure \ref{fig:GeVModel} on the left for a Lorentz factor of 300 and 800, and the upper limits given by the GRECO and ELOWEN analyses. The MRF of the decoupling model as a function of $\zeta_n\xi_N$ and $\Gamma$ can be seen in Figure \ref{fig:GeVModel} on the right.

%
 \subsection{MeV Neutrino Emission Models}


In addition to high-energy and quasi-thermal neutrino production, gamma-ray bursts (GRBs) hold promise as sites for thermal neutrino production through various mechanisms. One such scenario occurs when the explosion follows a core-collapse supernova, leading to the arrival of a large flux of MeV neutrinos before the shock breakout \citep{Kistler2013}. Another scenario involves the formation of an accretion disk around a black hole, known as neutrino-dominated accretion flows (NDAF). 
In NDAFs, the disk system becomes extremely hot and dense, causing photons to become trapped, while neutrinos can escape, carrying away gravitational energy from the black hole, cooling the system \citep{Liu2016}. The neutrinos from NDAFs can be detected at various stages, ranging from before the gamma-ray burst \citep{Wei2019, Liu2016} to the prompt gamma-ray phase \citep{Liu2016}. The diverse processes that give rise to these neutrinos make them intriguing messengers for probing the deeper and denser regions of GRBs, enabling the exploration of a wide parameter space, such as the black hole spin parameter or the accretion rate \citep{Liu2016}.
\smallskip

\begin{figure}[hbt!]
    \centering
    \includegraphics[width = 0.8\textwidth]{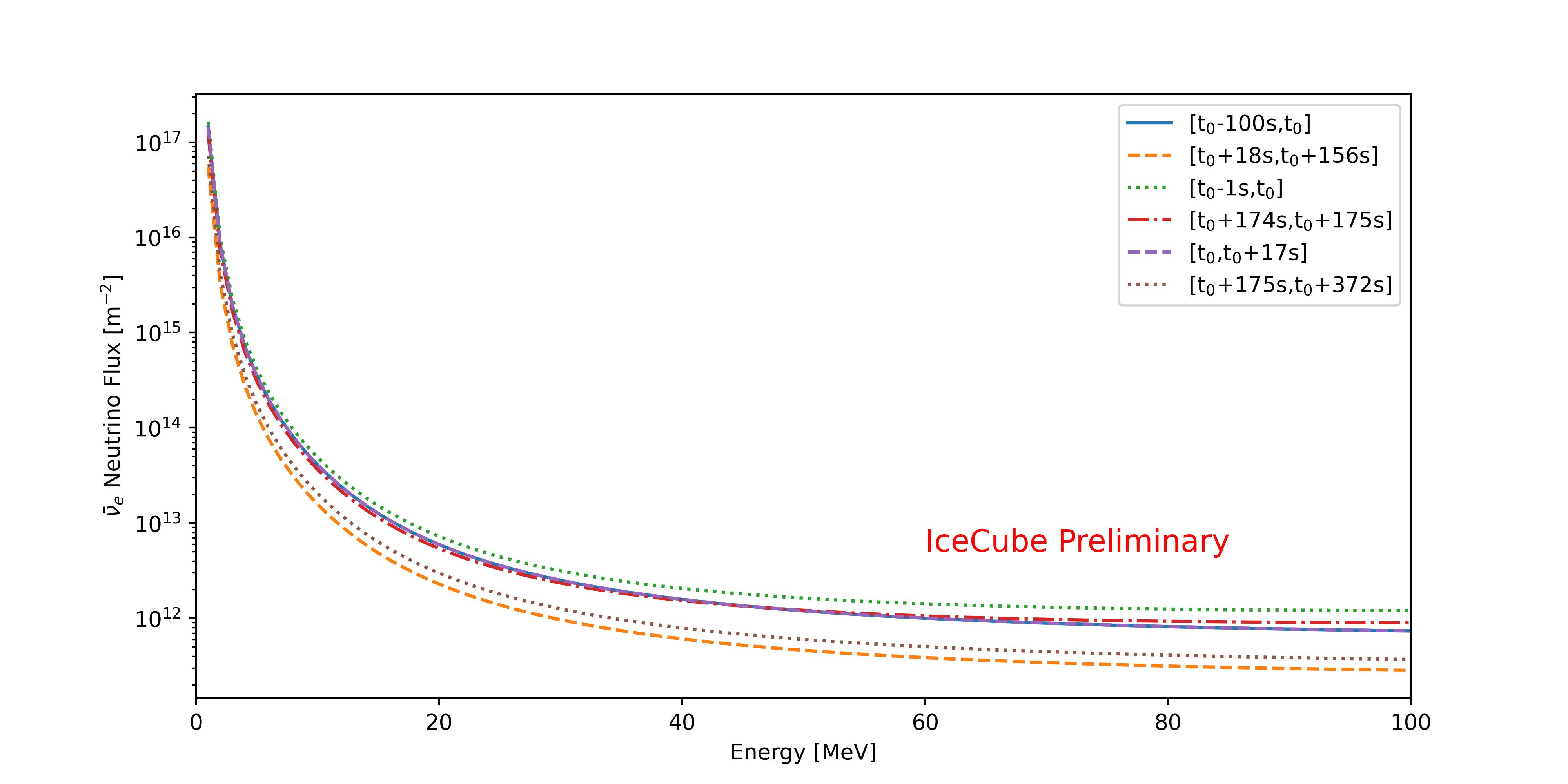}
    \caption{90\% C.L upper limits on the neutrino flux for the six time window searches used in the MeV analysis. These flux upper limits are calculated using a delta function in the given energy bin, where all of the energy is deposited in a single energy bin.}
    \label{fig:MeV_dirac_delta}
\end{figure}

Although IceCube has the capability to detect a burst of MeV neutrinos through an increased number of single photon hit rates, the considerable source distance (z=0.151 or $\sim$ 740 Mpc) poses a challenge in placing constraints on the parameter space for this specific GRB. Consequently, it is not currently feasible to test individual models based on the observed data. However, in this proceeding, we present a model-independent upper limit for this GRB utilizing a Supernova Test Routine for IceCube Analysis (ASTERIA) \citep{Griswold2020}.
\smallskip

In contrast to the previous paper \citep{IceCube:2023rhf}, which assumed a quasi-thermal spectrum with an average neutrino energy of $\Bar{E}_{\nu_e}=15$ MeV, we adopt a different approach in this proceeding. We assume a delta-function shape for the upper-limit total $\bar{\nu}_e$ luminosity with mono-energetic bins. Using this method (as described in \citep{TalkMeVConstraints}), we obtain model-independent upper limits (Fig. \ref{fig:MeV_dirac_delta}) on the $\bar{\nu}_e$ neutrino flux (1-100 MeV) for each time window described in \citep{IceCube:2023rhf}.

\section{Conclusion and prospects}\label{sec4}

With the search for neutrinos from GRB 221009A, upper limits on neutrino emission can be set in a wide energy range. We present the differential limits of the ELOWEN, GRECO and GFU analyses. With the upper limits, constraints can be set on different models predicting neutrino emission at varying locations and through varying processes. For the TeV neutrinos the ICMART and internal shock model were tested, which leads to improved limits compared to previous searches. In the GeV range, constraints can be put on the quasithermal decoupling and collision model by ELOWEN and GRECO, for different values of Lorentz factor and baryonic loading. Finally, for the MeV neutrinos, a model-independent upper limit for this GRB was calculated in the 6 different time windows.

Because of the electromagnetic brightness of GRB 221009A, we are able to place strong constraints on models of neutrino emission from this source. In addition, future improvements to the IceCube detector in different energy regimes will improve our sensitivity to neutrinos from GRBs. The IceCube-Upgrade will improve the detection of low-energy neutrinos, as well as the angular resolution of these neutrinos \cite{Ishihara:2019aao}. We are also improving the sensitivity of ELOWEN with the current detector setup \cite{TalkELOWEN}. Finally, the proposed extension of IceCube-Gen2 would significantly improve the sensitivity to high-energy neutrinos, where the goal is a sensitivity 5 times better than the current IceCube array \cite{IceCube-Gen2:2020qha}. Together, these developments will provide an important step towards detecting neutrinos originating from GRBs. 

\bibliographystyle{ICRC}
\bibliography{references}

%

\clearpage

\section*{Full Author List: IceCube Collaboration}

\scriptsize
\noindent
R. Abbasi$^{17}$,
M. Ackermann$^{63}$,
J. Adams$^{18}$,
S. K. Agarwalla$^{40,\: 64}$,
J. A. Aguilar$^{12}$,
M. Ahlers$^{22}$,
J.M. Alameddine$^{23}$,
N. M. Amin$^{44}$,
K. Andeen$^{42}$,
G. Anton$^{26}$
% ...
\\
\\
% ...
$^{12}$ Universit{\'e} Libre de Bruxelles, Science Faculty CP230, B-1050 Brussels, Belgium \\
% ...
$^{17}$ Department of Physics, Loyola University Chicago, Chicago, IL 60660, USA \\
$^{18}$ Dept. of Physics and Astronomy, University of Canterbury, Private Bag 4800, Christchurch, New Zealand \\
% ...
$^{22}$ Niels Bohr Institute, University of Copenhagen, DK-2100 Copenhagen, Denmark \\
$^{23}$ Dept. of Physics, TU Dortmund University, D-44221 Dortmund, Germany \\
% ...
$^{26}$ Erlangen Centre for Astroparticle Physics, Friedrich-Alexander-Universit{\"a}t Erlangen-N{\"u}rnberg, D-91058 Erlangen, Germany \\
% ...
$^{40}$ Dept. of Physics and Wisconsin IceCube Particle Astrophysics Center, University of Wisconsin{\textendash}Madison, Madison, WI 53706, USA \\
% ...
$^{42}$ Department of Physics, Marquette University, Milwaukee, WI, 53201, USA \\
% ...
$^{44}$ Bartol Research Institute and Dept. of Physics and Astronomy, University of Delaware, Newark, DE 19716, USA \\
% ...
$^{63}$ Deutsches Elektronen-Synchrotron DESY, Platanenallee 6, 15738 Zeuthen, Germany  \\
$^{64}$ Institute of Physics, Sachivalaya Marg, Sainik School Post, Bhubaneswar 751005, India \\
% ...

\subsection*{Acknowledgements}

\noindent
The authors gratefully acknowledge the support from the following agencies and institutions:
USA {\textendash} U.S. National Science Foundation-Office of Polar Programs,
U.S. National Science Foundation-Physics Division,
U.S. National Science Foundation-EPSCoR,
Wisconsin Alumni Research Foundation,
Center for High Throughput Computing (CHTC) at the University of Wisconsin{\textendash}Madison,
Open Science Grid (OSG),
Advanced Cyberinfrastructure Coordination Ecosystem: Services {\&} Support (ACCESS),
Frontera computing project at the Texas Advanced Computing Center,
U.S. Department of Energy-National Energy Research Scientific Computing Center,
Particle astrophysics research computing center at the University of Maryland,
Institute for Cyber-Enabled Research at Michigan State University,
and Astroparticle physics computational facility at Marquette University;
Belgium {\textendash} Funds for Scientific Research (FRS-FNRS and FWO),
FWO Odysseus and Big Science programmes,
and Belgian Federal Science Policy Office (Belspo);
Germany {\textendash} Bundesministerium f{\"u}r Bildung und Forschung (BMBF),
Deutsche Forschungsgemeinschaft (DFG),
Helmholtz Alliance for Astroparticle Physics (HAP),
Initiative and Networking Fund of the Helmholtz Association,
Deutsches Elektronen Synchrotron (DESY),
and High Performance Computing cluster of the RWTH Aachen;
Sweden {\textendash} Swedish Research Council,
Swedish Polar Research Secretariat,
Swedish National Infrastructure for Computing (SNIC),
and Knut and Alice Wallenberg Foundation;
European Union {\textendash} EGI Advanced Computing for research;
Australia {\textendash} Australian Research Council;
Canada {\textendash} Natural Sciences and Engineering Research Council of Canada,
Calcul Qu{\'e}bec, Compute Ontario, Canada Foundation for Innovation, WestGrid, and Compute Canada;
Denmark {\textendash} Villum Fonden, Carlsberg Foundation, and European Commission;
New Zealand {\textendash} Marsden Fund;
Japan {\textendash} Japan Society for Promotion of Science (JSPS)
and Institute for Global Prominent Research (IGPR) of Chiba University;
Korea {\textendash} National Research Foundation of Korea (NRF);
Switzerland {\textendash} Swiss National Science Foundation (SNSF);
United Kingdom {\textendash} Department of Physics, University of Oxford.

\end{document}